\documentclass[10pt]{PoS}
\usepackage[utf8]{inputenc}
\usepackage[T1]{fontenc}
\usepackage[UKenglish]{babel}
\usepackage{lmodern, xcolor, amsmath, graphicx, adjustbox}
\usepackage[numbers, elide, compress]{natbib}

\graphicspath{{fig/}}

\newcommand{\ee}{\ensuremath{e^+e^-}}
\newcommand{\bwbw}{\ensuremath{bW^+\bar bW^-}}
\newcommand{\eett}{\ensuremath{\ee\to t\,\bar t}}
\newcommand{\eebwbw}{\ensuremath{\ee\to\bwbw}}

\newcommand{\FDF}[1][]{\varphi^\dagger #1\!\overleftrightarrow{D}\!_\mu\varphi}
\newcommand{\FDFI}[1][]{\varphi^\dagger #1\!\overleftrightarrow{D}^I\!\!\!_\mu\:\varphi}

\let\Re\undefined
\let\Im\undefined
\DeclareMathOperator{\Re}{Re}
\DeclareMathOperator{\Im}{Im}

\newcommand{\infb}{\ensuremath{\,\text{fb}^{-1}}}
\newcommand{\inab}{\ensuremath{\,\text{ab}^{-1}}}
\newcommand{\gev}{\ensuremath{\,\text{GeV}}}
\newcommand{\tev}{\ensuremath{\,\text{TeV}}}

\DeclareMathAlphabet{\mathsfit}{\encodingdefault}{\sfdefault}{m}{sl}
\newcommand{\ges}[1]{\mathsfit{#1}}

\title{Precision constraints on the top-quark effective field theory at future lepton colliders\thanks{Based on work in collaboration with Martín Perelló, Marcel Vos and Cen Zhang.}}

\ShortTitle{Precision constraints on the top-quark effective field theory at future lepton colliders}

\author{Gauthier Durieux\\
        DESY, Notkestraße 85, D-22607 Hamburg, Germany\\
        E-mail: \email{gauthier.durieux@desy.de}
        }

\abstract{We examine the constraints that future lepton colliders would impose on the effective field theory describing modifications of top-quark interactions beyond the standard model, through measurements of the \eebwbw\ process. Statistically optimal observables are exploited to constrain simultaneously and efficiently all relevant operators. Their constraining power is sufficient for quadratic effective-field-theory contributions to have negligible impact on limits which are therefore basis independent. This is contrasted with the measurements of cross sections and forward-backward asymmetries. An overall measure of constraints strength, the global determinant parameter, is used to determine which run parameters impose the strongest restriction on the multidimensional effective-field-theory parameter space.
}

\FullConference{
XXV International Workshop on Deep-Inelastic Scattering and Related Subjects\\
3-7 April 2017\\
University of Birmingham, UK}

\begin{document}
\let\section\paragraph
\let\subsection\subparagraph

\section{Effective field theory}
The standard-model effective field theory (EFT) has the astonishing feature of parametrizing systematically the theory space in direct vicinity of the standard model (SM). Its applicability requires new physics to be heavier than the energies directly probed in the measurement considered. As a proper quantum field theory, it also allows for quantum corrections to be computed consistently.
We adopt such a description of top-quark interactions beyond the standard model, while most of the literature on this subject relies on anomalous vertices. This latter framework however suffers from several insufficiencies. It for instance, in general, allows for gauge-invariance violation in top electromagnetic couplings, misses four-fermion operators which can be generated at tree level by heavy mediators, or precludes the combination of constraints arising from various sectors, like the top and bottom sectors.
We rely on the so-called Warsaw basis of standard-model dimension-six operators \cite{Grzadkowski:2010es} and focus on the operators which interfere with standard-model \eebwbw\ amplitudes, at leading-order and in the massless-$b$ limit. Altogether, one then counts ten real degrees of freedom among which two violate CP. From the relevant two-quark and two-quark-two-lepton operators forming our effective Lagrangian $\mathcal{L}_\text{EFT} = \sum_i
\left(\frac{C_i}{\Lambda^2} O_i + \text{h.c.}\right)$,
\begin{gather}
\begin{array}{@{}rlcc@{}}
	O_{\varphi q}^1
		&\equiv \frac{y_t^2}{2}
		&\bar{\ges q}\gamma^\mu \ges q
		&\FDF[i]
	,\\
	O_{\varphi q}^3
		&\equiv \frac{y_t^2}{2}
		&\bar{\ges q}\tau^I\gamma^\mu \ges q
		&\FDFI[i]
	,\\
	O_{\varphi u}
		&\equiv \frac{y_t^2}{2}
		&\bar{\ges u}\gamma^\mu \ges u
		&\FDF[i]
	,\\
	O_{\varphi ud}
		&\equiv \frac{y_t^2}{2}
		&\bar{\ges u}\gamma^\mu \ges d
		&\varphi^T\!\epsilon\: iD_\mu\varphi
	,
\end{array}
\quad
\begin{array}{@{}rlcc@{}}
	O_{uG}
		&\equiv y_t g_s
		&\bar{\ges q}T^A\sigma^{\mu\nu} \ges u
		&\epsilon\varphi^* G_{\mu\nu}^A
	,\\
	O_{uW}
		&\equiv y_t g_W
		&\bar{\ges q}\tau^I\sigma^{\mu\nu} \ges u
		&\epsilon\varphi^* W_{\mu\nu}^I
	,\\
	O_{dW}
		&\equiv y_t g_W
		&\bar{\ges q}\tau^I\sigma^{\mu\nu} \ges d
		&\epsilon\varphi^* W_{\mu\nu}^I
	,\\
	O_{uB}
		&\equiv y_t g_Y
		&\bar{\ges q}\sigma^{\mu\nu} \ges u
		&\epsilon\varphi^* B_{\mu\nu}
	,
\end{array}
\quad
\begin{array}{@{}rlcc@{}}
	O_{u\varphi}
		&\equiv y_t^3 
		&\bar{\ges q} \ges u
		&\epsilon\varphi^* \; \varphi^\dagger\varphi
	,
\end{array}
\raisetag{.75cm}
\label{eq:op_2q}
\\
\begin{array}{rlcc}
	O_{lq}^1
		&\equiv
		&\bar{\ges q}\gamma_\mu \ges q
		&\bar{\ges l}\gamma^\mu \ges l
	,\\
	O_{lq}^3
		&\equiv
		&\bar{\ges q}\tau^I\gamma_\mu \ges q
		&\bar{\ges l}\tau^I\gamma^\mu \ges l
	,\\
	O_{lu}
		&\equiv
		&\bar{\ges u}\gamma_\mu \ges u
		&\bar{\ges l}\gamma^\mu \ges l
	,\\
	O_{eq}
		&\equiv
		&\bar{\ges q}\gamma_\mu \ges q
		&\bar{\ges e}\gamma^\mu \ges e
	,\\
	O_{eu}
		&\equiv
		&\bar{\ges u}\gamma_\mu \ges u
		&\bar{\ges e}\gamma^\mu \ges e
	,
\end{array}
\qquad
\begin{array}{rlccc}
	O_{lequ}^T
		&\equiv
		&\bar{\ges q}\sigma^{\mu\nu} \ges u
		&\epsilon
		&\bar{\ges l}\sigma_{\mu\nu} \ges e
	,
\end{array}
\qquad
\begin{array}{rlcc}
	O_{lequ}^S
		&\equiv
		&\bar{\ges q}\ges u
		&\epsilon\;
		\bar{\ges l}\ges e
	,\\
	O_{ledq}
		&\equiv
		&\bar{\ges d}\ges q
		&\bar{\ges l}\ges e
	,
\end{array}
\label{eq:op_2q2l}
\end{gather}
they are the ten combinations corresponding to:
\begin{equation}
\begin{array}{c}
	\begin{aligned}
		C_{lq}^A	&\equiv C_{lu} - (C_{lq}^1-C_{lq}^3),	\\
		C_{lq}^V	&\equiv C_{lu} + (C_{lq}^1-C_{lq}^3),
	\end{aligned}
	\qquad\quad
	\begin{aligned}
		C_{eq}^A	&\equiv C_{eu} - C_{eq}	,	\\
		C_{eq}^V	&\equiv C_{eu} + C_{eq}	,
	\end{aligned}
	\qquad\quad
	\begin{aligned}
		C_{\varphi q}^A	&\equiv C_{\varphi u} - (C_{\varphi q}^1-C_{\varphi q}^3),	\\
		C_{\varphi q}^V	&\equiv C_{\varphi u} + (C_{\varphi q}^1-C_{\varphi q}^3),
	\end{aligned}
\\[5mm]
	\begin{aligned}
	C^{R}_{uA}
		&= \Re\{ C_{uW} + C_{uB} \},\\
	C^{I}_{uA}
		&= \Im\{ C_{uW} + C_{uB} \},
	\end{aligned}
	\qquad\quad
	\begin{aligned}
	C^{R}_{uZ}
		&=\Re\{ c_W^2 C_{uW} - s_W^2 C_{uB} \}/s_Wc_W,\\
	C^{I}_{uZ}
		&=\Im\{ c_W^2 C_{uW} - s_W^2 C_{uB} \}/s_Wc_W,
	\end{aligned}
\end{array}
\label{eq:dof}
\end{equation}
where it is understood that the quark and lepton flavour indices are fixed to the third and first generation, respectively. The CKM matrix is approximated as unity, and the scale $\Lambda$ is conventionally set to $1\tev$.

\section{Observables and sensitivities}
\label{sec:obs}
To constrain globally this EFT parameter space, we introduce two sets of observables measured in a ILC-like benchmark scenario with runs at $500\gev$ and $1\tev$ centre-of-mass energies where, respectively, $500\infb$ and $1\inab$ of integrated luminosity are equally shared between $P(e^+,e^-)=(+0.3,-0.8)$ and $(-0.3,+0.8)$ beam polarization configurations.

\begin{figure}[tb]
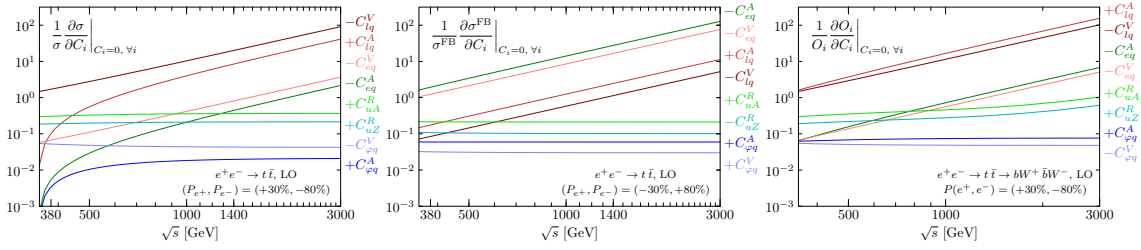
\centering
\includegraphics[width=.33\textwidth]{sensitivities_left_s.mps}%
\includegraphics[width=.33\textwidth]{sensitivities_right_a.mps}%
\includegraphics[width=.33\textwidth]{sensitivities_left_opt.mps}%
\vspace{-2mm}%
\caption{Sensitivities of the total (left), forward-backward-integrated (centre) cross sections, and statistically optimal observables (right) as functions of the centre-of-mass energy, for mostly left- (left, right) and mostly right-handed (centre) electron beam polarizations.}
\label{fig:sensitivities}
\vspace{-2mm}%
\end{figure}

\subsection{Cross sections and forward-backward asymmetries}
The first set of observables includes the total \bwbw\ cross sections and the $bW^+$ forward-backward asymmetries for each of these four runs. For a total measurement efficiency of $20\%$, these eight observables can be measured with a statistical uncertainty of about half a percent:
\begin{small}%
\begin{equation}%
\begin{array}{c@{\hspace{1cm}}cc@{\hspace{1cm}}cc}
 &\multicolumn{2}{c}{$500\infb\text{ at }500\gev$\hspace{1cm}}
 &\multicolumn{2}{c}{$1\inab\text{ at }1\tev$}
\\
P(e^+,e^-)=
 & (+0.3,-0.8) & (-0.3,+0.8) & (+0.3,-0.8) & (-0.3,+0.8)
\\[1mm]
\delta\sigma/\sigma=
& 0.0046	& 0.0065	& 0.0060	& 0.0082
\\
\delta A^\text{FB}=
 & 0.0043	& 0.0057	& 0.0051	& 0.0065
\end{array}.
\end{equation}%
\end{small}%
They are not linearly sensitive to the $C^I_{uZ}$ and $C^I_{uA}$ CP-violating coefficients but constitute a set sufficient to determine the eight remaining Wilson coefficients of \eqref{eq:dof}.

As can be seen in the left panel of Fig.\,\ref{fig:sensitivities}, the sensitivity of the total cross section to the axial-vector operator combinations $C_{lq}^A$, $C_{eq}^A$, $C_{\varphi q}^A$ suffers from a $(1-4m_t^2/s)^{1/2}$ suppression close to the top pair production threshold. The sensitivity to four four-fermion operator coefficients $C_{lq,eq}$ grows quadratically with energy as naively expected with dimension-six operators. On the contrary, it tends to a constant for the two $C_{\varphi q}$ since the two Higgs fields the corresponding operators contain condense to their vacuum expectation value to give rise to modifications of the SM $t\bar tZ$ coupling scaling as $v^2/\Lambda^2$. Following the same reasoning, a linear growth would have been expected for the $C_{uZ}$, $C_{uA}$ coefficients of dipole operators which contain one single Higgs field. A chirality flip required for their interferences with standard-model amplitudes however yields a $vm_t/\Lambda^2$ scaling. Interferences growing like $v\sqrt{s}/\Lambda^2$ can be recovered when the azimuthal helicity angles of the top decay products are not integrated over. These few features of the sensitivities are explicit in the \eett\ helicity amplitudes (for a $+-$ initial state), which have a 
\begin{equation}
\begin{array}{r@{\quad}l}
++:& \frac{2m_t}{\sqrt{s}} V + \sqrt{s}\: (D-\beta \tilde{D}),\\{}
--:& \frac{2m_t}{\sqrt{s}} V + \sqrt{s}\: (D+\beta \tilde{D}),
\end{array}
\qquad
\begin{array}{r@{\quad}l}
+-:& (V+\beta A) + 2m_t D,\\[.5mm]
-+:& (V-\beta A) + 2m_t D,
\end{array}
\label{eq:helicity_amps}
\end{equation}
schematic form where $V$, $A$, $D$, and $\tilde D$ respectively represent the contributions of the vector, axial-vector, magnetic, and electric dipole operators.
The central panel of Fig.\,\ref{fig:sensitivities} shows the sensitivity of the forward-backward-integrated cross section (such that $A^\text{FB}=\sigma^\text{FB}/\sigma$) to axial-vector operators is enhanced compared to that of the total cross section. This can be understood by realizing that $\sigma^\text{FB}\propto |+-|^2-|-+|^2$. The flip of beam polarization from the first panel to the second also inverts the hierarchy in sensitivities to the coefficients of four-fermion operators featuring left- ($C_{lq}$) and right-handed ($C_{eq}$) charged leptons.

\subsection{Statistically optimal observables}
Such observables are designed to yield the strongest global statistical constraints on a set of vanishing parameters appearing linearly in some differential distribution described by a model~\cite{Atwood:1991ka, *Davier:1992nw, *Diehl:1993br}. Assuming a
$
f(\Phi) = f_0(\Phi) + \sum_i C_i f_i(\Phi)
$
distribution where $\Phi$ specifies the phase space and given a sample of $n$ events, optimal observables are defined as the expectation values of $nf_i(\Phi)/f_0(\Phi)$ and are thus estimated by:
\vspace*{-3mm}
\begin{equation}
\bar O_i = \sum_{k=1}^n \frac{f_i(\Phi_k)}{f_0(\Phi_k)}.
\end{equation}
This approach is close to that of the matrix element method and should yield similar results under identical assumptions. In practice and in the context we are interested in, it can however be advantageous to define a discrete set of genuine observables and study, with traditional methods, how they are affected by higher-order corrections, detector effects, etc.\ (see also Ref.\,\cite{Gritsan:2016hjl} using a ``minimal set of determinants calculated as ratios of matrix elements'').

We use for $f(\Phi)$ the full five-dimensional $\eett\to\bwbw$ differential cross section depending on one top production angle, as well as one polar and one azimuthal angle characterizing the distribution of the decay products of each top. Analytical expressions are used for leading-order helicity amplitudes, in the vanishing $b$ mass limit and narrow-width approximation for the tops. The ones for instance found in Ref.\,\cite{Schmidt:1995mr} are complemented with contributions from four-fermion operators. Note these optimal observables depend on the beam energy and polarization. They are, in practice, not exactly optimal because the model used for their definitions describes real observations with a limited accuracy, because systematic uncertainties also play a role, and because the optimization is only performed in a linear approximation around the $C_i=0,\forall i$ point.

As seen in the right panel of Fig.\,\ref{fig:sensitivities}, since they use all the kinematic information available, the statistically optimal observables yield a sensitivity to dipole operator coefficients ($C_{uZ}^R$, $C_{uA}^R$) which mildly grows with energy in the range considered. Note however that optimal observables are not designed to maximize sensitivities, and that the definition of the sensitivity used does not make sense for the statistically optimal observables corresponding to CP-violating EFT parameters, as their expectation values vanish in the standard model (with CP-violating phase neglected, at tree level, in the narrow width approximation).

\begin{figure}[tb]
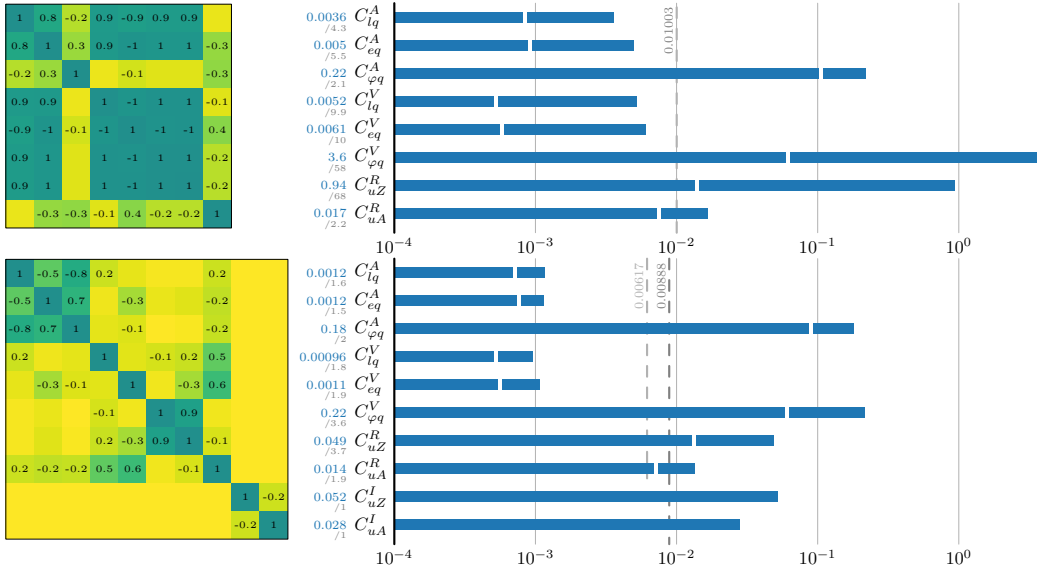
\centering
\scalebox{.9}{%
\begin{minipage}{\textwidth}
\includegraphics[scale=.825]{fit_def_benchmark_corr.mps}\hfill%
\raisebox{-3.5mm}{\includegraphics[scale=.75]{fit_def_benchmark.mps}}%
\\
\includegraphics[scale=.825]{fit_opt_benchmark_corr.mps}\hfill%
\raisebox{-3.5mm}{\includegraphics[scale=.75]{fit_opt_benchmark.mps}}%
\par%
\end{minipage}%
}
\caption{Global constraints deriving from the measurements of either cross section and forward-backward asymmetries (top) or optimal observables (bottom). Correlation matrices are displayed on the left. White marks indicate individual constraints and grey numbers, their ratio to marginalized constraints. Dashed vertical lines provide the average of constraint strengths in the form of global determinant parameters.}
\label{fig:fit}
\end{figure}

\begin{figure}[tb]
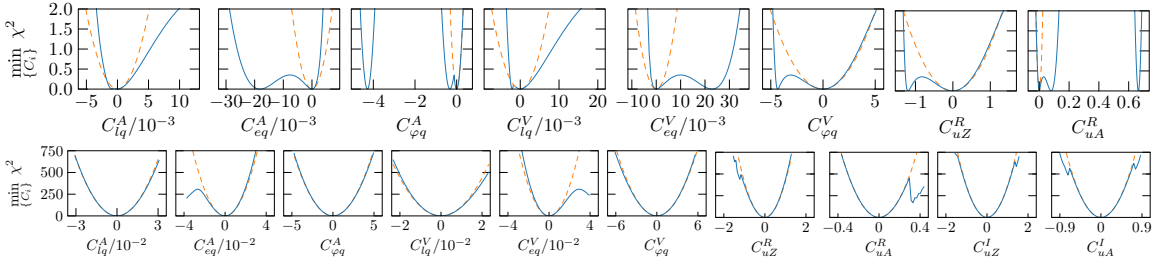

\adjustbox{max width=\textwidth}{%
\includegraphics{mchi_def_0.mps}%
\includegraphics{mchi_def_1.mps}%
\includegraphics{mchi_def_2.mps}%
\includegraphics{mchi_def_3.mps}%
\includegraphics{mchi_def_4.mps}%
\includegraphics{mchi_def_5.mps}%
\includegraphics{mchi_def_6.mps}%
\includegraphics{mchi_def_7.mps}}%
\\
\adjustbox{max width=\textwidth}{%
\includegraphics{mchi_opt_0.mps}%
\includegraphics{mchi_opt_1.mps}%
\includegraphics{mchi_opt_2.mps}%
\includegraphics{mchi_opt_3.mps}%
\includegraphics{mchi_opt_4.mps}%
\includegraphics{mchi_opt_5.mps}%
\includegraphics{mchi_opt_6.mps}%
\includegraphics{mchi_opt_7.mps}%
\includegraphics{mchi_opt_8.mps}%
\includegraphics{mchi_opt_9.mps}}%
\caption{Profiled chi-square's deriving from the measurements of cross sections and forward-backward asymmetries (top), and statistically optimal observables (bottom, note the unreasonably large $y$-axis scale), when only the linear dependence on dimension-six operator coefficients is accounted for (dashed) or when both linear and quadratic terms are included (solid).}
\label{fig:mchi}
\end{figure}

\begin{figure}[tb]
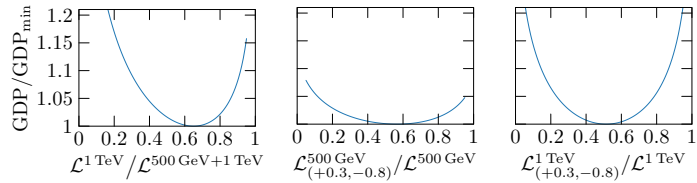
\centering
\adjustbox{width=.6\textwidth}{%
\includegraphics{gdp_opt_tev.mps}\quad%
\includegraphics{gdp_opt_x1.mps}\quad%
\includegraphics{gdp_opt_x2.mps}}%
\vspace{-2mm}
\caption{Variations of the GDP deriving from optimal observable measurements, as functions of the share of integrated luminosity spent at $500\gev$ and $1\tev$, and with two beam polarization configurations, $P(e^+,e^-)=(+0.3,-0.8)$ and $(-0.3,+0.8)$.}
\label{fig:gdp_opt}
\vspace{-3mm}
\end{figure}

\section{Global constraints}
For the ILC-like run scenario described above and an overall effective efficiency of $20\%$, Fig.\,\ref{fig:fit} shows the global statistical constraints deriving from the two sets of ideal measurements considered. A metric for the average constraint strength, the global determinant parameter or GDP~\cite{Durieux:2017rsg}, is defined from the determinant of $V$ the covariance matrix of the Gaussian fit to $N$ effective-field-theory parameters as $\text{GDP}\equiv \sqrt[2N]{\det V^{-1}}$. Interestingly, ratios of GDPs for different machines, run scenarios, or set of measurements are independent of rescalings and rotations in the EFT parameter space. They are thus operator basis independent. In terms of a GDP ratio, the constraints on CP-conserving parameters obtained from the measurements of statistically optimal observables are a factor $1.6$ times stronger than the ones obtained from cross-section and forward-backward asymmetry measurements.

The global constraints in Fig.\,\ref{fig:fit} are imposed after truncating the EFT expansion to the linear level. Some relatively loose constraints and large correlations obtained using cross-section and forward-backward asymmetry measurements however raise questions about the possible importance of quadratic $C_iC_j/\Lambda^4$ terms. When included, the top row of Fig.\,\ref{fig:mchi} shows that constraints are indeed significantly altered. On the contrary, the bounds derived from the measurements of statistically optimal observables remain unchanged (see bottom row of Fig.\,\ref{fig:mchi}). They thus have the much desired feature of being operator-basis independent, given that different dimension-six operator bases lead to inequivalent $C_iC_j/\Lambda^4$ contributions.

The GDP introduced above can also be employed to find the combination of run parameters leading to the strongest overall constraints. Note that different optimal run parameters could be obtained within specific models, or when a power counting is imposed. For a fixed integrated luminosity shared between runs at $500\gev$ and $1\tev$ centre-of-mass energies and between two beam polarizations, the optimal repartition leads to performances indistinguishable from our benchmark ILC-like run scenario. GDP variations around this minimum are displayed in Fig.\,\ref{fig:gdp_opt}.

\vspace*{-.5mm}
\section{Conclusions}
We studied the sensitivities of two sets of observables and the constraints future leptonic colliders would impose, through pair production, on the effective field theory of top-quark interactions. The power of statistically optimal observables yielding operator-basis-independent constraints has been demonstrated. The robustness of these observables against non-resonant contributions, beam structure, higher-order QCD corrections, realistic reconstruction and detector simulation is currently under investigation. Uses of the global determinant parameter to assess the global strengthening of constraints with different set of measurements and run parameters have been illustrated.

\vspace*{-.5mm}
\section{Acknowledgements}
The author thanks Marc Montull and François Le Diberder for discussions respectively on the inequivalence of quadratic contributions in different dimension-six operator bases, and on the similarity between statistically optimal observables and matrix element methods.

\vspace*{-.9mm}
\setlength{\bibsep}{.5mm}
\renewcommand{\bibfont}{\raggedright}
\bibliographystyle{apsrev4-1_title.bst}
\bibliography{proceedings.bib}
\end{document}